\documentclass[preprint,onecolumn,nofootinbib]{revtex4}
\pdfoutput=1
\usepackage[colorlinks=true,linkcolor=blue,urlcolor=blue,filecolor=black,citecolor=red,pdfstartview=FitV,pdftitle={},pdfsubject={},pdfkeywords={},pdfpagemode=None,bookmarksopen=true]{hyperref}
\usepackage{graphicx}
\usepackage{amsmath}
\usepackage{amsfonts}
\usepackage{amssymb,ulem}
\usepackage{dcolumn,booktabs}%
\usepackage{color,xcolor}%
\usepackage{CJK}
\usepackage{subfigure}

\usepackage{dcolumn}
\usepackage{float}
\usepackage{multirow}
\usepackage{dcolumn,booktabs}
\usepackage{enumitem}
\usepackage{amssymb}
\newcommand{\f}{\begin{equation}}
\newcommand{\ff}{\end{equation}}
\newcommand{\fa}{\begin{eqnarray}}
\newcommand{\ffa}{\end{eqnarray}}

\begin{document}
\title{Imprints of quantum gravity effects on gravitational waves: a comparative study using extreme mass-ratio inspirals}

\author{Ruo-Ting Chen$^{1}$}
\thanks{ruotingchen@163.com}
\author{Guoyang Fu$^{1}$}
\thanks{FuguoyangEDU@163.com}
\author{Dan Zhang$^{1}$}
\thanks{danzhanglnk@163.com}
\author{Jian-Pin Wu$^{1}$}
\thanks{jianpinwu@yzu.edu.cn, corresponding author} 
\affiliation{
  $^1$ Center for Gravitation and Cosmology, College of Physical
  Science and Technology, Yangzhou University, Yangzhou 225009,
  China}

\begin{abstract}

Within a generally covariant Hamiltonian framework of loop quantum gravity (LQG), two black hole models parameterized by a quantum correction $\zeta$ have recently been constructed. Using extreme mass-ratio inspirals (EMRIs) as high-precision probes, we investigate the imprints of this LQG deformation in the surrounding spacetime. Waveforms generated via an improved augmented analytic kludge (AAK) model in both LQG black hole backgrounds and in Schwarzschild spacetime are compared through a faithfulness analysis. This allows us to quantify the detectability of the deviation with LISA and to derive constraints on 
$\zeta$ based on a detection threshold. We find that the first LQG black hole model produces significantly stronger signatures in EMRI signals than the second, making its quantum gravity effects more accessible to future space-borne gravitational-wave detection.

\end{abstract}

\maketitle
\tableofcontents

\section{Introduction}

Since its formulation, Einstein’s general relativity (GR) has provided an exceptionally successful framework for gravitation and the large-scale structure of spacetime \cite{Einstein:1915ca, Einstein:1916vd}. To date, it has withstood virtually all existing experimental and observational tests, from solar system dynamics \cite{Will:2014kxa, GRAVITY:2020gka} to the recent detections of gravitational waves (GWs) \cite{LIGOScientific:2016aoc, LIGOScientific:2016lio, LIGOScientific:2016sjg, LIGOScientific:2019fpa, LIGOScientific:2025rid} and the images of M 87 or Sgr A$^{*}$ \cite{EventHorizonTelescope:2019dse, EventHorizonTelescope:2019ths, EventHorizonTelescope:2022xnr, EventHorizonTelescope:2022xqj}, the latter two also confirmed the existence of black holes (BHs).
Despite these successes, GR continues to encounter several fundamental theoretical challenges. A prominent example is the singularity theorems, formulated by Penrose and Hawking in the 1970s \cite{Penrose:1964wq, Penrose:1969pc, Hawking:1970zqf}, which demonstrate the generic inevitable occurrence of spacetime singularities within GR. The existence of singularities leads to the non-extendibility of the spacetime manifold and to the geodesic incompleteness of timelike or null trajectories.
In such extreme regimes, the failure of the classical description of gravity and the breakdown of known physical laws motivate the search for a more fundamental framework. This pursuit has led to the development of quantum gravity theories, such as loop quantum gravity (LQG) \cite{rovelli2004quantum, Ashtekar:2004eh}, which aim to provide a consistent description of spacetime and gravitation at the Planck scale, and have become an important focus of contemporary research in cosmology and BH physics.

Given that spacetime covariance is a cornerstone of GR, its consistent implementation has been a primary theoretical pursuit in the development of the loop quantum gravity black hole (LQG-BH) models, with notable progress recently achieved (see Refs. \cite{Alonso-Bardaji:2021yls, Alonso-Bardaji:2022ear}). In particular, building on the general scheme proposed in \cite{Bojowald:2011aa}, C. Zhang et al. constructed the corresponding effective metric and Hamiltonian constraint while explicitly respecting covariance \cite{Zhang:2024khj}. By applying two distinct polymerization schemes to the Dirac observable, i.e. the effective mass, they obtained two respective forms of the Hamiltonian constraint. Following the procedure outlined in \cite{Cafaro:2024vrw}, the resulting dynamics were solved, yielding two classes of static, spherically symmetric BH solutions. 
In the first class, the metric components satisfy $-g_{00}=g^{11}$, leading to a Reissner-Nordström (RN)-like structure. The second class features a metric whose temporal component coincides with that of the Schwarzschild solution, while the radial component $g_{11}$ contains a nontrivial correction function. In both cases, the quantum gravity effects are characterized by the parameter $\zeta$, which removes the classical singularity through a bounce region, resulting in a symmetric spacetime connecting to a white hole on the opposite side.
These two families of LQG-BH solutions have subsequently been studied extensively across various physical contexts. For detailed investigations into their properties, perturbations, observational signatures, and related phenomenological implications, we refer the reader to Refs. \cite{Yang:2025ufs, Konoplya:2024lch, Skvortsova:2024msa, Liu:2024soc, Liu:2024wal, Du:2024ujg, Xamidov:2025oqx, Umarov:2025wzm, Chen:2025aqh, Ban:2024qsa, Lin:2024beb, Heidari:2024bkm, Wang:2024iwt, Shu:2024tut, Chen:2025ifv, Du:2025kcx, Liu:2025hcx, Al-Badawi:2025yqu, Zhang:2025ccx, Sahlmann:2025fde, Liu:2025iby}.

As indicated in previous studies \cite{Yang:2025ufs, Konoplya:2024lch, Skvortsova:2024msa, Liu:2024soc, Liu:2024wal, Du:2024ujg, Xamidov:2025oqx, Umarov:2025wzm, Chen:2025aqh, Ban:2024qsa, Lin:2024beb, Heidari:2024bkm, Wang:2024iwt, Shu:2024tut, Chen:2025ifv, Du:2025kcx, Liu:2025hcx, Al-Badawi:2025yqu, Zhang:2025ccx, Sahlmann:2025fde, Liu:2025iby}, these quantum gravity effects may exist outside the horizon and could, in principle, be detectable through strong-field GW observations.
GWs provide one of the most precise probes to study the detailed structure of BH spacetimes and their surrounding astrophysical environment \cite{Nair:2022xfm, Mitra:2023sny, Krishnendu:2025byo, Battista:2021rlh, Battista:2022hmv, DeFalco:2023djo, DeFalco:2024ojf}.
Among various GW sources, extreme mass-ratio inspirals (EMRIs) stand out as high-precision laboratories for testing fundamental gravity, owing to their extreme mass ratios, high expected event rates, and large signal-to-noise ratios (SNRs) \cite{Amaro-Seoane:2007osp, Cardenas-Avendano:2024mqp}. A typical EMRI system consists of a stellar-mass compact secondary object orbiting and gradually inspiraling toward a supermassive BH, emitting GWs at the same time, with mass ratios generally ranging from $10^{-7}$ to $10^{-4}$. By the late stages of inspiral, the secondary can complete on the order of $10^{4}$ to $10^{5}$ orbital cycles. This long-duration, cumulative evolution allows EMRI waveforms to encode exceptionally rich information about the strong-field spacetime, making them one of the most promising sources for future space-borne GW detectors such as LISA \cite{Babak:2017tow, LISA:2022kgy}, Taiji \cite{Gong:2021gvw}, and TianQin \cite{TianQin:2020hid}. At the same time, the intricate nature of these signals poses substantial challenges for accurate waveform modeling and data analysis. 
High fidelity EMRI waveform construction requires accounting for gravitational self-force effects \cite{Barack:2018yvs}, which have so far been developed up to second order in Schwarzschild spacetime \cite{Wardell:2021fyy}. Such calculations, however, are extremely complex and computationally expensive. Consequently, semi-classical kludge approximations — including the analytic kludge (AK) \cite{Barack:2003fp}, numerical kludge (NK) \cite{Babak:2006uv}, and augmented analytic kludge (AAK) \cite{Chua:2017ujo, Katz:2021yft} — are widely used as efficient waveform templates. These models reproduce the main physical features of realistic EMRI signals while greatly reducing computational cost.

EMRIs have been applied to a wide range of research topics. They have been used, for instance, to test modified gravity theories, including those involving dark matter \cite{Dai:2023cft, Zhang:2024ugv} or scalar fields \cite{Guo:2022euk, Barsanti:2022vvl, Zhang:2022rfr, Zi:2025lio}, and to probe possible deviations from GR \cite{Kumar:2024utz, Ahmed:2025azu}. They also serve as standard sirens for constraining the cosmological parameters \cite{Laghi:2021icm, Laghi:2021pqk}. In addition, they have inspired the development of specialized subfields, such as studies of dirty EMRIs, which focus on inspirals occurring in environments with accretion and matter around galactic-center BHs \cite{Lyu:2024gnk}, and binary EMRIs, in which a binary system inspirals as a whole toward a supermassive BH \cite{Chen:2018axp}. 
More recently, EMRI systems have also been proposed as probes of quantum gravity effects \cite{Fu:2024cfk, Liu:2024qci, Zi:2024jla, Yang:2024cnd, Yang:2025esa, Gong:2025mne, Ahmed:2025shr}. These works suggest that GW signals from EMRIs could carry subtle imprints of quantum gravity, and that EMRI observations can place tighter constraints on LQG parameters than those obtained from weak-field solar system experiments \cite{Liu:2022qiz, Chen:2023bao, Ai:2025myf}, highlighting EMRIs as powerful tools for exploring the quantum aspects of gravity.

Following this framework, we investigate quantum gravity effects on the external spacetime by employing EMRI system around two aforementioned types of covariant LQG-BHs.
The corresponding EMRI signals are generated using an AAK waveform model improved with the \texttt{FastEMRIWaveforms} (FEW) scheme \cite{Katz:2021yft}. We then evaluate the detectability of such quantum gravity effects with LISA and derive constraints on the fundamental correction parameter $\zeta$. This study aims to quantify the deviations of LQG-BHs from their GR counterparts and to compare the differences between the two LQG spacetimes, thereby providing potential waveform templates for future observations and contributing an essential step in theoretical testing.

The paper is organized as follows. In Sec. \ref{bhsolution}, we briefly introduce the spacetime properties of the two LQG-corrected BH solutions and present the general equations of motion for a secondary body orbiting either primary BH in an EMRI setup. In Sec. \ref{OF}, we formulate the adiabatic evolution of the orbit under gravitational radiation, derive the associated energy and angular momentum fluxes, and present the resulting time evolution equations. The generation of EMRI gravitational waveforms and their detectability by the LISA mission are discussed in Sec. \ref{Wave}.  Finally, Sec. \ref{conclusion} summarizes our findings and discusses future prospects.

Throughout this paper we adhere to the $\left(-, +, +, +\right)$ signature for the metrics. The Planck units, i.e. $G=c=\hbar=1$ is adopted in theoretical calculations. When generating waveform data from EMRI system, we revert to the international system of units. Latin letters $\left\{i, j, k\right\}$ are spatial components range over $1, 2, 3$, while Greek indices range over $0, 1, 2, 3$. The analysis is performed in Schwarzschild coordinate system $x^{\mu}=\left(x^{0}, x^{1}, x^{2} , x^{3}\right)\equiv \left(t, r, \theta, \phi\right)$.

\section{Motions of massive particles in effective LQG-BHs}\label{bhsolution}

In this section, we briefly review the spacetime geometrical properties of the two LQG-BH backgrounds derived from the condition of general covariance \cite{Zhang:2024khj}. We then derive the universal equations of motion for a secondary compact object orbiting these central primary LQG-BHs in the context of an EMRI, laying the groundwork for the subsequent inspiral evolution analysis.

\subsection{Effective LQG-BHs}

Under the requirement of general covariance, the authors obtained an effective Hamiltonian constraint formalism and, via specialized polymerizations of Dirac observables, yielded two effective BH metrics parametrized by a quantum correction.
The spherically symmetric exteriors of these two LQG-BHs can be described by the following general expression \cite{Zhang:2024khj}:
\begin{eqnarray}
\mathrm{d}s^{2}=-f\left(r\right) \mathrm{d}t^{2}+\frac{1}{g\left(r\right) f\left(r\right)} \mathrm{d}r^{2}+r^2 \left(\mathrm{d}\theta^2 +\sin^2\theta\thinspace \mathrm{d}\phi^2 \right) . \label{equ:II.A-(1)}
\end{eqnarray}
The metric functions for the two types are given separately. For the type-I BH:
\begin{eqnarray}
f_{\mathrm{I}}\left(r\right)=1-\frac{2 M}{r}+\frac{M^2 \tilde{\zeta}^2}{r^2} \left(1-\frac{2 M}{r}\right)^2, \quad
g_{\mathrm{I}}\left(r\right)=1 .
\label{typeI_BH}
\end{eqnarray}
For the type‑II BH:
\begin{eqnarray}
f_{\mathrm{II}}\left(r\right)=1-\frac{2 M}{r}  ,\quad 
g_{\mathrm{II}}\left(r\right)=1+\frac{M^2 \tilde{\zeta}^2}{r^2} \left(1-\frac{2 M}{r}\right).
\label{typeII_BH}
\end{eqnarray}
In both cases $M$ denotes the Arnowitt-Deser-Misner mass of the spacetime. 
A dimensionless parameter $\tilde{\zeta}$ is introduced to encode the leading quantum gravity corrections, which are tied to the Planck length $l_\mathrm{P}$. It is defined as:
\begin{eqnarray}
\tilde{\zeta} \equiv \frac{\gamma \sqrt{4\sqrt{3}\pi \gamma l_{\mathrm{P}}^{2}}}{M} ,
\end{eqnarray}
where $\gamma$ is the Barbero-Immirzi parameter which relates to the BH entropy \cite{Domagala:2004jt}, and the Planck length is set to unity, $l_{\rm{P}}=1$, henceforth. Notably, the quantum gravity parameter $\tilde{\zeta}$ is proportional to $\sqrt{\hbar}$, encoding the minimal length scale expected in LQG. Consequently, $\tilde{\zeta}^2$ defines the minimum area gap in LQG, a key feature that reflects the discrete structure of spacetime at the Planck scale. Without loss of generality, we neglect the tilde notation for the expression simplification.
In the following analysis, we treat $\gamma$ as a variable, thereby enabling a direct investigation of the characteristics of $\zeta$.
Additionally, unless stated otherwise, subscripts ``I" and ``II" denote the first and second types of the LQG-BH model, respectively. In the absence of subscripts, the results are applicable to both cases. 

Prior to advancing, we offer insights into the global geometric properties of these effective LQG-based BHs. For the type-I metric \eqref{typeI_BH}, the RN-like formalism gives two horizons where the outer one is the classical BH horizon locates at $r_{+}=2M$, whereas the inner horizon $r_{-}$ is located at:
\begin{eqnarray}
r_{-} = \frac{M \zeta^{4/3}}{\left(-27 + 3\sqrt{81 + 3 \thinspace \zeta^{2}}\right)^{1/3}}-\frac{M \zeta^{2/3} \left(-27 + 3\sqrt{81 + 3 \thinspace \zeta^{2}}\right)^{1/3}}{3} \,.
\label{r_minus}
\end{eqnarray} 
Based on the analysis in Ref. \cite{Zhang:2024khj}, the type-I LQG-BH spacetime shares the same causal structure as the LQG-BH metrics derived from the gravitational collapse model proposed in Refs. \cite{Munch:2021oqn, Yang:2022btw}. This structure features a transition region between the bounce surface $r_{\mathrm{b}}$ and the inner horizon $r_{-}$, which connects to another white hole spacetime.

\begin {figure}[h]
    \centering
    \begin {minipage} {0.55\textwidth}
        \centering
        \includegraphics[
   width = \linewidth] {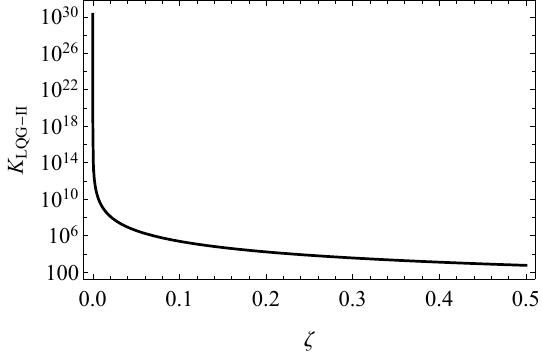}
    \end {minipage}
    \caption {Relation between $K_{\mathrm{LQG-II}}$ and the LQG parameter $\zeta$ at the bounce radius $r_{\mathrm{b}}$
.}
    \label {K_zeta}
\end {figure}

For the type-II solution, the bounce surface coincides with the inner horizon, $r_{\mathrm{b}}=r_{-}$, serving as the non-singular bridge between the BH and a white hole spacetime (see the supplementary material of Ref. \cite{Zhang:2024khj} for more detailed discussions).
We analyze the nature of this surface using the Kretschmann scalar, which  diverges at a true singularity but remains finite at a coordinate singularity. Denoting the Kretschmann scalar of this LQG spacetime as $K_{\mathrm{LQG-II}}$, the computed expression is given below:
\begin{eqnarray}
K_{\mathrm{LQG-II}} &=& \frac{48 M^{2}}{r^{6}}-\frac{16 M^{3}\left(21 M^{2}-14 M r+2 r^{2}\right) \zeta^{2}}{r^{9}} \nonumber\\[3mm]
&+& \frac{4 M^{4}\left(201 M^{4}-274 M^{3} r+139 M^{2} r^{2}-32 M r^{3}+3 r^{4}\right) \zeta^{4}}{r^{12}} .  
\label{Kre_2}
\end{eqnarray} 
Substituting the Eq. \eqref{r_minus} into the above expression, we can numerically plot $K_{\mathrm{LQG-II}}$ against the correction parameter $\zeta$ (see Fig. \ref{K_zeta}). 
It is evident that quantum gravity effects render $K_{\mathrm{LQG-II}}$ finite, and its value decreases monotonically with increasing $\zeta$. This confirms that $r_{\mathrm{b}}$ is merely a coordinate singularity. Moreover, in the limit $\zeta\to 0$, Eq. \eqref{Kre_2} reduces to the classical result (the leading order term), thereby recovering the GR scenario.

\subsection{Geodesic for massive particle}

In the EMRI system, we can approximate the secondary compact object as a test particle of mass $m$. Its dynamical evolution in the LQG-BH spacetime is derived from the following Lagrangian:
\begin{eqnarray}
\mathcal{L}=\frac{1}{2} m g_{\mu \nu} \dot{x}^{\mu} \dot{x}^{\nu} ,
\end{eqnarray}
where the overdot ``.'' denotes the first derivative of ${x}^{\mu}$ with respect to the affine parameter $\tau$ along the timelike geodesics. 
The canonical momentum of the particle conjugate to $x^{\mu}$ is defined as:
\begin{eqnarray}
p_{\mu } \equiv \frac{\partial \mathcal{L} }{\partial \dot{x}^{\mu }  }= m g_{\mu \nu } \dot{x}^{\nu }  .
\label{p_mu}
\end{eqnarray}
From this definition, its four components in the given metric can be written explicitly as follows:
\begin{eqnarray}
&&p_{t} = m \left(-f\left(r\right)\right) \dot{t} , \label{p_t} \\[3mm]
&&p_{r} = m \frac{1}{g\left(r\right) f\left(r\right)} \dot{r}, \\[3mm]
&&p_{\theta} = m r^{2} \dot{\theta}, \\[3mm]
&&p_{\phi} = m r^{2} \sin ^{2} \theta \thinspace \dot{\phi} . \label{p_phi}
\end{eqnarray}

Then, we obtain the Hamiltonian by utilizing the Legendre transformation: $\mathcal{H} = p_{\mu} \dot{x}^{\mu }-\mathcal{L}= \left(1/2m\right) g^{\mu \nu} p_{\mu} p_{\nu}$. From the canonical phase space, the Poisson brackets yield two conserved quantities:
\begin{eqnarray}
\dot p_{t} = \left \{ p_{t},\mathcal{H} \right \} = 0,  \quad
\dot{p}_{\phi} = \left\{p_{\phi}, \mathcal{H} \right \}= 0 ,
\end{eqnarray}
which correspond to the energy $E$ and angular momentum $L$ of the massive particle respectively. 
Through the Eqs. \eqref{p_t} and \eqref{p_phi}, we can obtain the equations of $t$ and $\phi$: 
\begin{eqnarray}
\dot{t} &=& \frac{p_{t}}{m \left(-f\left(r\right)\right)} = \frac{E}{m f\left(r\right)} , \\[3mm]
\dot{\phi} &=& \frac{p_{\phi}}{m r^{2} \sin ^{2} \theta} = \frac{L}{m r^{2} \sin ^{2} \theta} .
\end{eqnarray}
Utilizing the above equations and the four-velocity normalization $g^{\mu \nu} p_{\mu} p_{\nu} = -m^{2}$, we can formulate the following expression:
\begin{eqnarray}
-\frac{{E}^{2}}{f\left(r\right)} + g\left(r\right) f\left(r\right) \thinspace p_{r}^{2}+\frac{p_{\theta}^{2}}{r^{2}} +\frac{{L}^{2}}{r^{2} \sin ^{2} \theta}
= -m^{2} .
\label{reduced_H}
\end{eqnarray}

By introducing the Carter constant $\mathcal{C}$ \cite{Carter:1968rr} to separate the components $r$ and $\theta$, we can identify a third constant of motion as follows:
\begin{eqnarray}
\mathcal{C} &=& p_{\theta}^{2}+\frac{L^{2}}{\sin ^{2} \theta} = -m^{2} r^{2} + \frac{r^{2} E^{2}}{f\left(r\right)}-r^{2} g\left(r\right) f\left(r\right) \thinspace p_{r}^{2} ,
\label{sep_C}
\end{eqnarray}
Subsequently, the equations for $r$ and $\theta$ are derived as:
\begin{eqnarray}
&& \dot{\theta}=\frac{{\mathcal{C}}}{m^{2} r^{4}}-\frac{L^{2}}{m^{2} r^{4} \sin ^{2} \theta }, \label{theta_dot} \\[3mm]
&& \dot{r} = \frac{{E}^{2} g\left(r\right)}{m^{2}} - \frac{\mathcal{C} \thinspace g\left(r\right) f\left(r\right)}{m^{2} r^{2}} - g\left(r\right) f\left(r\right) .
\end{eqnarray}
For the simplification, we mainly focus on the evolution of motion in the equatorial plane, where $\theta={\pi}/{2}$, $\dot{\theta}=0$, then the Carter constant in Eq. \eqref{theta_dot} reduces to $\mathcal{C} = L^{2}$. As a result, the geodesic motion of the massive particle is governed by:
\begin{eqnarray}
&& \dot{t} =\frac{E}{m f\left(r\right)} , \label{eq:tdot}  \\[3mm]
&& \dot{\phi} = \frac{L}{m r^{2} } ,  \label{eq:phidot}  \\[3mm]
&& \dot{r} = \frac{{E}^{2} g\left(r\right)}{m^{2}} - \frac{L^{2} g\left(r\right) f\left(r\right)}{m^{2} r^{2}} - g\left(r\right) f\left(r\right) .
\label{eq:rdot}
\end{eqnarray}
Once we have the geodesic equations available, we can delve deeper into exploring the impact of quantum gravity on the evolution of the secondary body through EMRI system.

\section{Equatorial orbit evolution}\label{OF}

In this section, we first establish the motion of the secondary body in an eccentric Keplerian equatorial orbit around these two LQG-BHs, extracting two fundamental frequencies that serve as key inputs for constructing the kludge waveform. We then incorporate radiation reaction into the dynamics to drive the inspiral, treating the system within the adiabatic and weak-field approximations. With the orbit-averaged energy and angular momentum fluxes, we finally examine how quantum gravity modifies the evolution relative to GR, focusing on its impact on the time-dependent orbital parameters and phases evolution.

\subsection{Eccentric motion and fundamental orbital frequencies}\label{FOF}

Following the standard procedure, we assume the secondary compact object takes the eccentric motion, described by the classical Kepler equation. We transform the coordinates from $\left(t, r, \theta, \phi \right)$ to $\left(t, X, \theta, \phi \right)$, where the radial coordinate $r$ is parameterized by the true anomaly $X$ as:
\begin{eqnarray}
r\left( X \right) = \frac{M p}{1+e \cos X}  .
\label{eq:r_chi}
\end{eqnarray}
Here $M$ is the mass of the primary body in EMRI, the symbols $p$ and $e$ denote the dimensionless semi-latus rectum and eccentricity of the orbit, respectively. In this context $e$ ranges in $[0,1)$. Within one complete period, the true anomaly parameter $X$ varies from $0$ to $2\pi$,  leading to two extreme points locate at periapsis $r_{\mathrm{p}}$ and apoapsis $r_{\mathrm{a}}$, which can be expressed as:
\begin{eqnarray}
r_{\mathrm{p}} = \frac{M p}{1+e}  , \quad  
r_{\mathrm{a}} = \frac{M p}{1-e} .
\label{eq:rarp}
\end{eqnarray}

For the bound orbits, we have the condition $\dot{r}|_{r_{\mathrm{p}}, r_{\mathrm{a}}}=0$. 
Imposing this condition yields general expressions for the energy $E$ and angular momentum $L$, which can be written schematically as:
\begin{eqnarray}
&&E^2 = \frac{m^{2}\left({r_\mathrm{a}}^{2}-{r_\mathrm{p}}^{2}\right) f\left({r_\mathrm{a}}\right) f\left({r_\mathrm{p}}\right)}{-{r_\mathrm{p}}^{2} f\left({r_\mathrm{a}}\right) + {r_\mathrm{a}}^{2} f\left({r_\mathrm{p}}\right)} ,  \label{eq_E2}  \\[3mm] 
&&L^2 = \frac{m^{2} {r_\mathrm{a}}^{2} {r_\mathrm{p}}^{2}\left(f\left({r_\mathrm{a}}\right)- f\left({r_\mathrm{p}}\right)\right)}{-{r_\mathrm{p}}^{2} f\left({r_\mathrm{a}}\right) + {r_\mathrm{a}}^{2} f\left({r_\mathrm{p}}\right)} .
\label{eq_L2}
\end{eqnarray}
Combining the above equations with Eqs. \eqref{eq:rdot} and \eqref{eq:rarp}, we obtain the explicit forms of $E_{\rm{I}}$ and $L_{\rm{I}}$ in the type-I LQG-BH background as
\begin{eqnarray}
&&{E_{\rm{I}}}^2 = \frac{m^{2}\left[ \left(p-2 \right)^2 - 4e^2\right]\left[ p^{3}+2(-1+e)^{3} \zeta^{2}+(-1+e)^{2} p \zeta^{2}\right]\left[ -p^{3}+2(1+e)^{3} \zeta^{2}-(1+e)^{2} p \zeta^{2}\right]}{\left(-3-e^{2}+ p\right) p^{4}\left[p^{3}-4\left(1+3 e^{2}\right) \zeta^{2}+2\left(1+e^{2}\right) p \zeta^{2}\right]} ,\label{eq:E_1}   \\[3mm]
&&{L_{\rm{I}}}^2 = \frac{m^{2} M^{2} p^{2}\left[p^{3}-8\left(1+e^{2}\right) \zeta^{2}+2\left(3+e^{2}\right) p \zeta^{2}-p^{2} \zeta^{2}\right]}{\left(-3-e^{2}+p\right)\left[p^{3}-4\left(1+3 e^{2}\right) \zeta^{2}+2\left(1+e^{2}\right) p \zeta^{2}\right]}    ,
\label{eq:L_1}
\end{eqnarray}
as well as those in the type-II LQG-BH background:
\begin{eqnarray}
&&{E_{\rm{II}}}^2 = \frac{m^2 \left[ -4 e^2 + \left(-2 + p \right)^2 \right]}{p \left(-3 - e^2 + p \right)} ,\label{eq:E_2}  \\[3mm]
&&{L_{\rm{II}}}^2 = \frac{m^2 M^2 p^2}{-3 - e^2 + p}    .
\label{eq:L_2}
\end{eqnarray}
It is noteworthy that the energy $E_{\rm{II}}$ and angular momentum $L_{\rm{II}}$ do not receive quantum gravity corrections and thus coincide with their Schwarzschild counterparts. This follows directly from the general expressions given in Eqs. \eqref{eq:E_2} and \eqref{eq:L_2} as they depend only on $f(r)$, a function that receives no quantum corrections in the BH-II background. In contrast, the BH-I metric \eqref{typeI_BH} explicitly incorporates quantum gravity effects through its $f(r)$. As expected, in the limit $\zeta\to 0$, the expressions for ${E_{\rm{I}}}^2$ and ${L_{\rm{I}}}^2$ reduce to their Schwarzschild counterparts.

In the EMRI scenario, the semi-latus rectum $p$ evolves by decreasing until the secondary object plunges. Here we analyze its critical value at the final stage. 
The equations of motion \eqref{eq:rdot} yield the following effective potential for the radial motion:
\begin{eqnarray}
V_{\mathrm{eff}}= \frac{E^2}{m^2}- \frac{\left[E^2 r^2-\left(L^2+m^2 r^2\right)f\left(r\right)\right] g\left(r\right)}{m^2 r^2}  .
\end{eqnarray}
For the marginally bound orbits, the conditions $V_{\mathrm{eff}}=E^2/m^2$ and $\partial_r V_{\mathrm{eff}}|_{r_{\mathrm{p}}}=0$ must be satisfied. These lead to two real root solutions for each of the two types of LQG-BHs, which are given by:
\begin{eqnarray}
p_1&=& 6+2e, \label{eq:p_1}  \\[3mm]
p_2&=& \frac{(1+{e}) \zeta^{2 / 3}\left[\left(9+\sqrt{3} \sqrt{27+\zeta^{2}}\right)^{2 / 3}-3^{1 / 3} \zeta^{2 / 3}\right]}{3^{2 / 3}\left(9+\sqrt{3} \sqrt{27+\zeta^{2}}\right)^{1 / 3}} . 
\label{eq:p_2}
\end{eqnarray}
Since the equation $\partial_r V_{\mathrm{eff}}=0$ for metric-I \eqref{typeI_BH} constitutes a higher-order polynomial that cannot be solved analytically, we performed a series expansion in the parameter $\zeta$ up to second order in the above calculation. 

The first solution for $p$ (Eq. \eqref{eq:p_1}) is constrained to the range $p\in \left(6+2e, \infty \right)$. This corresponds to a periastron radius $r_{\rm{p}}>{\left(6+2e \right)M}/\left(1+e\right)$, which implies $p>6$ and $r_{\rm{p}}>4M$ \cite{Cutler:1994pb, Sundararajan:2008bw}. The orbit becomes a plunging trajectory once the boundary $p=6+2e$ is crossed, making it the natural cutoff for our subsequent analysis.
The second solution for $p$ (Eq. \eqref{eq:p_2}) is beyond our study, as it is a consequence of the transition region involving quantum gravity and resides within the BH, thus we do not consider it further here.

Since we already set $\theta=\pi/2$ (the equatorial plane), the polar angle frequency $\Omega_{\theta}$ can out of consideration. The orbital evolution is thus governed by the two fundamental frequencies $\Omega_{r}$ and $\Omega_{\phi}$, which correspond to the radial $r\left(X\right)$ and azimuthal $\phi$, respectively. 

As the secondary mass completes one orbital period from the initial time $t=0$ to an arbitrary time $t=t_{0}$, the transformed coordinate $X$, obtained by mapping the radial coordinate $r$, ranges over the interval $[0, 2\pi]$.
Consequently, the corresponding variation of the coordinate time is expressed as:
\begin{eqnarray}
T_r = \int_{0}^{t_0}\mathrm{d}t = \int_{0}^{2\pi}\frac{\mathrm{d}t}{\mathrm{d}X}\mathrm{d}X  , 
\label{eq:T_r}
\end{eqnarray}
while the associated change in the azimuthal angle is denoted by:
\begin{eqnarray}
\Delta\phi_r = \int_{0}^{\phi_0}\mathrm{d}\phi =  \int_{0}^{2\pi}\frac{\mathrm{d}\phi}{\mathrm{d}X}\mathrm{d}X . 
\end{eqnarray}
It is emphasized that the chain rule was employed in the aforementioned derivations. 
We can therefore define the two fundamental frequencies $\Omega_{r}$ and $\Omega_{\phi}$ over each period as follows:
\begin{eqnarray}
\Omega_r &=& \frac{2\pi}{T_r} , \\ [3mm]
\Omega_{\phi} &=& \frac{\Delta{\phi_r}}{T_r} . 
\label{eq:Omega_phi}
\end{eqnarray}

We now focus on orbits that start in the weak-field region, i.e., $p\gg6$. By applying the Eqs. \eqref{eq:tdot} and \eqref{eq:phidot}, together with the LQG-BH functions \eqref{typeI_BH} and \eqref{typeII_BH} to the expressions above, we obtain the frequency-domain equations for metric-I and -II as expansions in powers of $p^{-1}$ as follows:
\begin{eqnarray}
{\Omega_{\phi}}_{\mathrm{I}} &=& \frac{\left(1-e^{2}\right)^{3 / 2}}{M p^{3 / 2}} + \frac{\left(1-e^{2}\right)^{3 / 2}\left(6e^{2}-\zeta^{2} \right)}{2 M p^{5 / 2}}+ \mathcal{O}\left(p^{-7/2}\right) , \label{eq:Omega_phi_1} \\[3mm]
{\Omega_r}_{\mathrm{I}} &=& \frac{\left(1-{e}^{2}\right)^{3 / 2}}{M p^{3 / 2}} - \frac{3 \left(1-{e}^{2}\right)^{5 / 2}}{M p^{5 / 2}} - \frac{3\left(1-{e}^{2}\right)^{5 / 2}\left(-2+6{e}^{2}+5 \sqrt{1-{e}^{2}}-\zeta^{2}\sqrt{1-{e}^{2}}\right)}{2 M p^{7 / 2}}  \nonumber\\[3mm]
&+& \mathcal{O}\left(p^{-9/2}\right)  , \label{eq:Omega_r_1}  \\[3mm]
{\Omega_{\phi}}_{\mathrm{II}} &=& \frac{\left(1-e^{2}\right)^{3 / 2}}{M p^{3 / 2}}+\frac{3 e^{2}\left(1-e^{2}\right)^{3 / 2}}{M p^{5 / 2}} + \frac{\left(1-e^{2}\right)^{3 / 2}}{4 M p^{7 / 2}}\left\{36 e^{4}+2\left(-1+\sqrt{1-e^{2}}\right)\left(-15+\zeta^{2}\right) \right.  \nonumber\\[3mm]
&-& \left. e^{2}\left[9-30 \sqrt{1-e^{2}}+\left(1+2 \sqrt{1-e^{2}}\right) \zeta^{2}\right]\right\} + \mathcal{O}\left(p^{-9/2}\right)  ,   \label{eq:Omega_phi_2} \\[3mm]
{\Omega_r}_{\mathrm{II}} &=& \frac{\left(1-{e}^{2}\right)^{3 / 2}}{M p^{3 / 2}} - \frac{3 \left(1-{e}^{2}\right)^{5 / 2}}{M p^{5 / 2}} - \frac{\left(1-{e}^{2}\right)^{5 / 2}\left(-6+18{e}^{2}+15 \sqrt{1-{e}^{2}}-\zeta^{2}\sqrt{1-{e}^{2}}\right)}{2 M p^{7 / 2}}  \nonumber\\[3mm]
&+& \mathcal{O}\left(p^{-9/2}\right) .
\label{eq:Omega_r_2}
\end{eqnarray}
Clearly, quantum gravity effects manifest as corrections to the frequency components, with the classical Schwarzschild limit recovered at $\zeta=0$. For practical computations, we employ higher-order expansions to achieve greater accuracy in the orbital evolution, the explicit form of which are not presented here.

\subsection{Fluxes and evolution}\label{evolution}

Building upon the preceding discussion, we now introduce gravitational radiation to drive the inspiral of bound orbits in LQG-BH backgrounds. Given that the radiation reaction timescale is much longer than the orbital period, we adopt the adiabatic approximation in our calculations \cite{Hughes:2005qb, Isoyama:2021jjd}. This allows us to treat the inspiral as a sequence of geodesics. The evolution between successive geodesics is determined by the emitted gravitational radiation, which is characterized by the energy and angular momentum fluxes.

In the weak-field limit, we employ the quadrupole formula, yielding the following equations for the rates of energy and angular momentum loss \cite{Peters:1963ux, Maggiore:2007ulw}:
\begin{eqnarray}
&&\frac{\mathrm{d}E}{\mathrm{d}t} = \frac{1}{5} \left \langle \frac{\mathrm{d}^3 \mathcal{Q}_{ij}}{\mathrm{d}t^3}\frac{\mathrm{d}^3 \mathcal{Q}_{ij}}{\mathrm{d}t^3} \right \rangle  ,
\label{eq:dE}  \\[3mm]
&&\frac{\mathrm{d}L_{i}}{\mathrm{d}t} = \frac{2}{5} \epsilon^{ijk} \left \langle \frac{\mathrm{d}^2 \mathcal{Q}_{jl}}{\mathrm{d}t^2}\frac{\mathrm{d}^3 \mathcal{Q}_{kl}}{\mathrm{d}t^3} \right \rangle ,
\label{eq:dL}
\end{eqnarray}
where the angle bracket $\left \langle \ \right \rangle$ denotes an average over one orbital period, and $\epsilon^{ijk}$ represents the Levi-Civita symbol. The symbol $\mathcal{Q}$ denotes the quadrupole moment, defined as $\mathcal{Q}^{ij} \equiv \mathcal{M}^{ij}-\left(1/3\right)\delta^{ij}\mathcal{M}_{kk}$ via the mass moments $\mathcal{M}^{ij}=\mu x^{i}x^{j}$. The reduced mass $\mu \equiv mM/\left(m+M\right)$ can be approximated by the secondary mass $m$ in the extreme mass-ratio limit, and the vector $x^{i}$ denotes the position of the secondary body relative to the primary, which we express in Cartesian coordinates as $x^{i}=\left( r\cos \phi, r\sin \phi, 0\right)$. 
Accordingly, through formulae \eqref{eq:dE} and \eqref{eq:dL}, we obtain the orbit-averaged energy flux and angular momentum flux described by $\left\{p, e\right\}$ in the first type LQG-BH background as: 
\begin{eqnarray}
\left \langle \frac{\mathrm{d}E}{\mathrm{d}t} \right \rangle_{\mathrm{I}} &=& \frac{(1-e^2)^{3/2}(96+292e^2+37e^4)m^2}{15M^2p^5} \nonumber\\[3mm]
&+& \frac{\left(1-e^{2}\right)^{3 / 2}\left[53 e^{6}-96 \zeta^{2}+e^{2}\left(176-416 \zeta^{2}\right)-6 e^{4}\left(-75+22 \zeta^{2}\right)\right]m^2}{5 M^2 p^{6}}+\mathcal{O}(p^{-7}) ,  \label{eq:adE_1}     \\[3mm]
\left \langle \frac{\mathrm{d}L}{\mathrm{d}t} \right \rangle_{\mathrm{I}} &=& \frac{4(1-e^2)^{3/2}(8+7e^2)m^2}{5Mp^{7/2}}\nonumber \\[3mm]
&-& \frac{2\left(1-e^{2}\right)^{3 / 2}\left[40 \zeta^{2}+2 e^{4}\left(-27+\zeta^{2}\right)+e^{2}\left(-76+63 \zeta^{2}\right)\right] m^2}{5 M p^{9 / 2}}+\mathcal{O}\left(p^{-11/2}\right)  .
\label{eq:adL_1}
\end{eqnarray}
and those in the metric-II background as:
\begin{eqnarray}
\left\langle \frac{\mathrm{d}E}{\mathrm{d}t} \right\rangle_{\mathrm{II}} &=& \frac{(1-e^2)^{3/2}(96+292e^2+37e^4)m^2}{15M^2p^5} + \frac{{e}^{2}\left(1- {e}^{2}\right)^{3 / 2}\left(176+450{e}^{2}+53{e}^{4}\right) m^2}{5 M^2 {p}^{6}} \nonumber\\[3mm]
&+& \frac{m^2}{60 M^2 {p}^{7}}\left(1-{e}^{2}\right)^{3 / 2}\left\{1908{e}^{8}+192\left(-1+\sqrt{1-{e}^{2}}\right)\left(-15+\zeta^{2}\right) \right. \nonumber \\[3mm]
&+& \left.8{e}^{2}\left[1923-735 \sqrt{1-{e}^{2}}+\left(-181+49 \sqrt{1-{e}^{2}}\right) \zeta^{2}\right] \right. \nonumber \\[3mm]
&+& \left.{e}^{6}\left[18427+1110 \sqrt{1-{e}^{2}}-\left(61+74 \sqrt{1-{e}^{2}}\right) \zeta^{2}\right] \right. \nonumber \\[3mm]
&-& \left. 6{e}^{4}\left[-5\left(434+255 \sqrt{1-{e}^{2}}\right)+\left(182+85 \sqrt{1-{e}^{2}}\right) \zeta^{2}\right]\right\} + \mathcal{O}\left(p^{-8}\right) , \label{eq:adE_2}   \\[3mm]
\left\langle \frac{\mathrm{d}L}{\mathrm{d}t} \right\rangle_{\mathrm{II}} &=&\frac{4(1-e^2)^{3/2}(8+7e^2)m^2}{5Mp^{7/2}} + \frac{4 e^{2}\left(1-e^{2}\right)^{3 / 2}\left(38+27 e^{2}\right)m^2}{5 M p^{9 / 2}} \nonumber \\[3mm]
&+& \frac{m^2}{10 M p^{11 / 2}}\left(1-e^{2}\right)^{3 / 2}\left\{675 e^{6}+32\left(-1+\sqrt{1-e^{2}}\right)\left(-15+\zeta^{2}\right) \right. \nonumber \\[3mm]
&-& \left. 4 e^{2}\left[-3\left(83+5 \sqrt{1-e^{2}}\right)+\left(31+\sqrt{1-e^{2}}\right) \zeta^{2}\right] \right. \nonumber \\[3mm]
&+& \left. e^{4}\left[957+420 \sqrt{1-e^{2}}-\left(33+28 \sqrt{1-e^{2}}\right) \zeta^{2}\right]\right\} + \mathcal{O}\left(p^{-13/2}\right) .
\label{eq:adL_2}
\end{eqnarray}
The chain rule was again applied in the above derivative calculations. 
It is observed that quantum gravity effects do not manifest at the leading order but enter at higher post-Newtonian (PN) orders. Furthermore, the parameter $\zeta$ enters at the subleading order in the fluxes for LQG-BH I but at the next-to-subleading order for LQG-BH II. When these quantum effects vanish, the flux formulae revert to the general relativistic counterparts, thereby explicitly quantifying the deviation from GR.

Having obtained the energy and angular momentum fluxes carried away by GWs per unit time, we proceed to compute the orbital evolution including gravitational radiation. Under the adiabatic approximation, the continuous loss of energy and angular momentum from the source can be identified with the averaged fluxes given by Eqs. \eqref{eq:adE_1} to \eqref{eq:adL_2}, namely, 
\begin{eqnarray}
\left\langle \frac{\mathrm{d}E}{\mathrm{d}t} \right\rangle_{\mathrm{GW}} &=& -\left\langle \frac{\mathrm{d}E}{\mathrm{d}t} \right\rangle = -\mu \dot{E}   ,  \label{eq:dE_GW} \\[3mm]
\left\langle \frac{\mathrm{d}L}{\mathrm{d}t} \right\rangle_{\mathrm{GW}} &=& -\left\langle \frac{\mathrm{d}L}{\mathrm{d}t} \right\rangle = -\mu \dot{L} ,
\label{eq:dL_GW}
\end{eqnarray}
where the minus sign reflects the balance between the orbital fluxes lost by the secondary and the outward flux of gravitational radiation to infinity. 
When expressed using the orbital parameters $p$ and $e$, Eqs. \eqref{eq:dE_GW} and \eqref{eq:dL_GW} can be reorganized as follows:
\begin{eqnarray}
-\left\langle \frac{\mathrm{d}E}{\mathrm{d}t} \right\rangle_{\mathrm{GW}} &=& \mu \frac{\partial E}{\partial p}\frac{\mathrm{d}p}{\mathrm{d}t} + \mu \frac{\partial E}{\partial e}\frac{\mathrm{d}e}{\mathrm{d}t}, \\[3mm]
-\left\langle \frac{\mathrm{d}L}{\mathrm{d}t} \right\rangle_{\mathrm{GW}} &=& \mu \frac{\partial L}{\partial p}\frac{\mathrm{d}p}{\mathrm{d}t} + \mu \frac{\partial L}{\partial e}\frac{\mathrm{d}e}{\mathrm{d}t}.
\end{eqnarray}
We recast the above equations and yield the time evolution rates of $\left\{p,e\right\}$ as:
\begin{eqnarray}
\mu \frac{\mathrm{d}p}{\mathrm{d}t} = \left(\frac{\partial E}{\partial e}{\left\langle \frac{\mathrm{d}L}{\mathrm{d}t} \right\rangle_{\mathrm{GW}}}-\frac{\partial L}{\partial e}{\left\langle \frac{\mathrm{d}E}{\mathrm{d}t} \right\rangle_{\mathrm{GW}}}\right)/\left(\frac{\partial L}{\partial e}\frac{\partial E}{\partial p}-\frac{\partial E}{\partial e}\frac{\partial L}{\partial p}\right) ,
\label{eq:dp}   \\[3mm]
\mu \frac{\mathrm{d}e}{\mathrm{d}t} = \left(\frac{\partial E}{\partial p}{\left\langle \frac{\mathrm{d}L}{\mathrm{d}t} \right\rangle_{\mathrm{GW}}}-\frac{\partial L}{\partial p}{\left\langle \frac{\mathrm{d}E}{\mathrm{d}t} \right\rangle_{\mathrm{GW}}}\right)/\left(\frac{\partial L}{\partial p}\frac{\partial E}{\partial e}-\frac{\partial L}{\partial e}\frac{\partial E}{\partial p}\right) .
\label{eq:de}
\end{eqnarray}
Here, we take the metric II as a representative example. Substituting Eqs. \eqref{eq:E_2} \eqref{eq:L_2} \eqref{eq:adE_2} \eqref{eq:adL_2} into Eqs. \eqref{eq:dp} and \eqref{eq:de}, one can achieve the following functional forms of $\left\{\dot{p}, \dot{e}\right\}_{\mathrm{II}}$:
\begin{eqnarray}
\frac{\mathrm{d}p}{\mathrm{d}t}_{\mathrm{II}} &=& -\frac{8 \left(1- {e}^{2}\right)^{3 / 2}\left(8+7 {e}^{2}\right)m}{5 M{p}^{3}}-\frac{2 \left(1- {e}^{2}\right)^{3 / 2}\left(144+326 {e}^{2}+245 {e}^{4}\right)m}{15 M{p}^{4}}  \nonumber \\[3mm]
&-& \frac{\left(1-{e}^{2}\right)^{3 / 2}m}{15 M{p}^{5}} \left\{1399 {e}^{6}+24\left[159-60 \sqrt{1-{e}^{2}}+4\left(-1+\sqrt{1-{e}^{2}}\right) \zeta^{2}\right]  \right. \nonumber\\[3mm]
&+& \left. {e}^{2}\left[3091+180 \sqrt{1-{e}^{2}}-12\left(31+\sqrt{1-{e}^{2}}\right) \zeta^{2}\right]  \right. \nonumber \\[3mm]
&+& \left. {e}^{4}\left[4\left(296+315 \sqrt{1-{e}^{2}}\right)-3\left(33+28 \sqrt{1-{e}^{2}}\right) \zeta^{2}\right]\right\} +\mathcal{O}\left({p}^{-11 / 2}\right)   ,  \label{eq:dp_2}   \\[3mm]
\frac{\mathrm{d}e}{\mathrm{d}t}_{\mathrm{II}} &=& -\frac{{e}\left(1-{e}^{2}\right)^{3 / 2}\left(304+121 {e}^{2}\right)m}{15 M{p}^{4}}-\frac{{e}\left(1-{e}^{2}\right)^{3 / 2}\left(1280+2097 {e}^{2}+697 {e}^{4}\right)m}{30 M{p}^{5}} \nonumber \\[3mm]
&-& \frac{{e}\left(1-{e}^{2}\right)^{3 / 2}m}{120 M{p}^{6}} \left\{7615 {e}^{6}+16\left[3087-1140 \sqrt{1-{e}^{2}}+4\left(-28+19 \sqrt{1-{e}^{2}}\right) \zeta^{2}\right] \right. \nonumber \\[3mm]
&+& \left. {e}^{2}\left[5\left(4711+2196 \sqrt{1-{e}^{2}}\right)-12\left(273+61 \sqrt{1-{e}^{2}}\right) \zeta^{2}\right] \right. \nonumber\\[3mm]
&+&\left. {e}^{4}\left[22676+7260 \sqrt{1-{e}^{2}}-2\left(259+242 \sqrt{1-{e}^{2}}\right) \zeta^{2}\right]\right\}+\mathcal{O}\left({p}^{-13 / 2} \right) . 
\label{eq:de_2}
\end{eqnarray}
It should be noted that the exact expressions are intricate. Therefore, the expansions above are provided primarily to present our main results clearly. 
As shown, the dominant negative sign on the right-hand side indicates that the orbital parameters $\left\{p,e\right\}$ decrease with time — behavior consistent with the inspiral of the secondary body toward the central BH. Quantum gravity corrections enter through higher-order terms and consequently modify the orbital evolution. The corresponding expressions $\left\{\dot{p}, \dot{e}\right\}_{\mathrm{I}}$ for the LQG-BH I background can be derived following the same procedure.

To illustrate the influence of the quantum parameter on the orbital dynamics within the two LQG-BH spacetimes, we numerically solve for the time evolution of $p\left(t\right)$ and $e\left(t\right)$ using the $\texttt{NDSolve}$ routine in $\mathit{Mathematica}$, with initial conditions $\left\{{p}_{0}, {e}_{0}\right\}$. According to the discussion in subsection~\ref{FOF}, the separatrix of the orbital evolution in EMRI is located at $p=6+2e$. To maintain numerical stability, we terminate the integration when the orbit reaches $p_{\mathrm{stop}}=6+2e+0.1$, i.e., 0.1 away from the separatrix.
The initial semi-latus rectum is fixed at $p_{0}=10$, providing at least one year of stable GW evolution within the interval $p\in \left[p_{\mathrm{stop}},\ p_{0}\right]$, and enabling a clear assessment of the parameter $\zeta$'s impacts on the waveform properties. For the initial eccentricity $e_{0}$, we consider values of 0.01 and 0.1. This choice is motivated by our strategy of ensuring that the orbital evolution remains sufficiently far from the circular bound orbit, corresponding to low-eccentricity elliptical orbits that retain moderate ellipticity without circularizing completely during the evolution. Under this setup, we concentrate on the GW emission characteristics of low-eccentricity elliptical orbits.

\begin {figure}[h]
    \centering
     \vspace{0.35cm}
      \includegraphics[width = 1.03\linewidth] {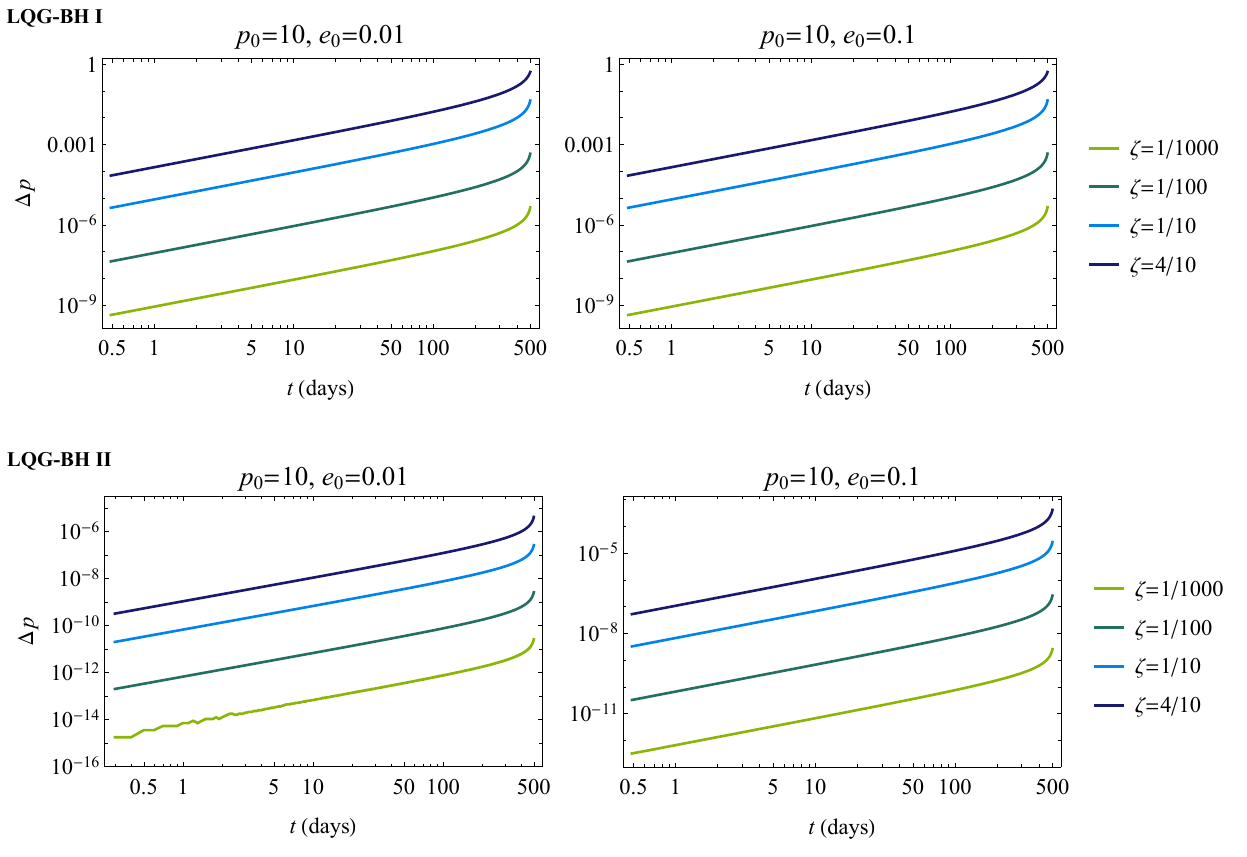}
    \caption {Deviation in the time evolution of the semi-latus rectum $\Delta p=p_{\mathrm{I, II}}-p_{\mathrm{Sch}}$ for different values of the quantum gravity parameter $\zeta$.
}
    \label {fig_dp}
\end {figure}

\begin {figure}[h]
    \centering
     \vspace{0.35cm}
      \includegraphics[width = 1.03\linewidth] {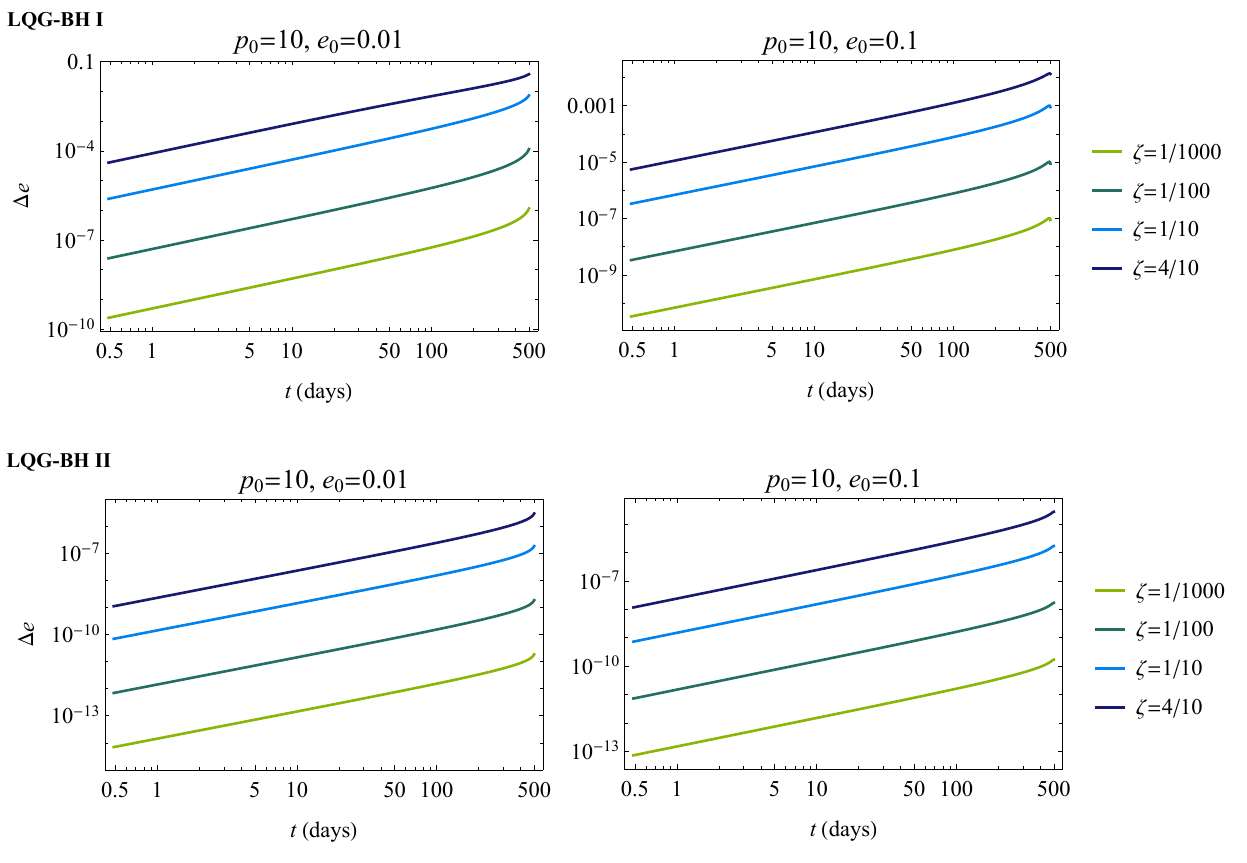}
    \caption {Deviation in the time evolution of the eccentricity $\Delta e=e_{\mathrm{I, II}}-e_{\mathrm{Sch}}$ for different values of the quantum gravity parameter $\zeta$.
}
    \label {fig_de}
\end {figure}

As the correction associated with the quantum parameter $\zeta$ arises only at higher orders, we separately compute the time evolution of the orbital parameter differences between the Schwarzschild spacetime (i.e., $\zeta = 0$) and the two LQG-BH spacetimes for various choice of $\zeta$, the corresponding results are presented on a log-log scale in Figs. \ref{fig_dp} and \ref{fig_de}. For the calculations, we adopt masses of $m=10 M_{\odot}$ for the secondary body and $M=10^{6} M_{\odot}$ for the central object, with $M_{\odot}$ denoting the solar mass. The numerical results show that the orbital deviations $\Delta p$ and $\Delta e$ for both types of LQG-BHs exhibit a monotonic increase over time, with their magnitudes growing as $\zeta$ increases. This behavior indicates that quantum gravity effects induce a cumulative impact on orbital evolution, thereby offering potential observability over long timescales. Moreover, a comparison between the two LQG-BH models reveals that for a given $\zeta$,
the orbital parameter deviations in the LQG-BH I spacetime are more pronounced than those in LQG-BH II. However, the evolution is only weakly affected by the choice of initial conditions. In other words, the orbital parameter differences in LQG-BH I are more sensitive to variations in $\zeta$ but less sensitive to the initial configuration.

\begin {figure}[h]
    \centering
     \vspace{0.35cm}
      \includegraphics[width = 1.03\linewidth] {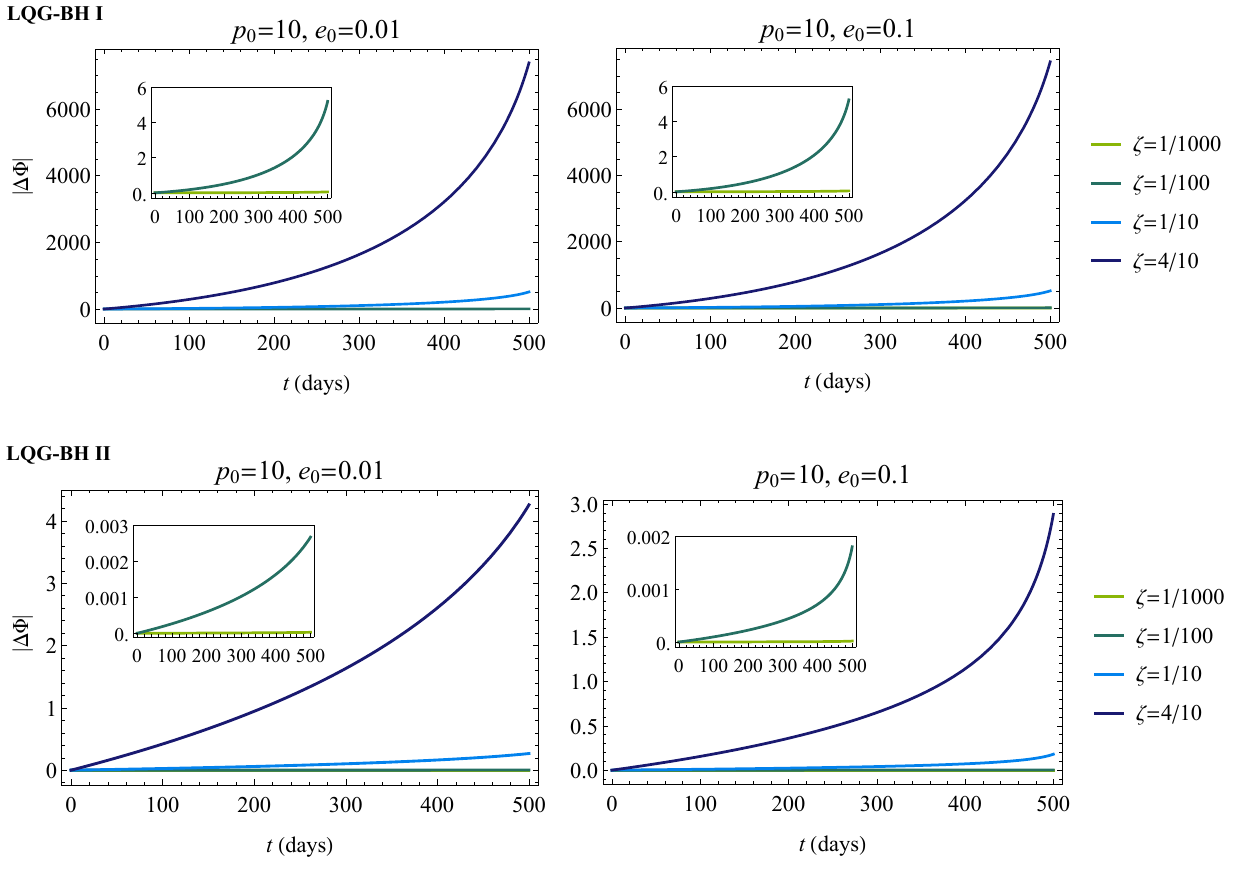}
    \caption {EMRI orbital dephasing $\left | \Delta \Phi \right |$ versus evolution time for different values of $\zeta$, where the inset presents enlarged views for $\zeta=1/1000$ and $\zeta = 1/100$.
}
    \label{fig_dPhi}
\end {figure}

We now proceed to analyze the orbital phases in the EMRI system. During the inspiral evolutions, there are two phases change with the fundamental azimuthal frequency $\Omega_{\phi}$ and radial frequency $\Omega_{r}$, we denote them as $\Phi_{\phi}$ and $\Phi_{r}$ respectively. Their average evolution rates are determined by Eqs. \eqref{eq:Omega_phi_1}--\eqref{eq:Omega_r_2} as follows:
\begin{eqnarray}\label{phases}
\frac{\mathrm{d}\Phi_{\phi,r}}{\mathrm{d}t}=
\frac{1}{T_r}\int_{0}^{2\pi}\Omega_{\phi,r}\left(p\left(t\right), e\left(t\right)\right) \frac{\mathrm{d}t}{\mathrm{d}X}\mathrm{d}X\,.
\end{eqnarray}
Evidently, the phase $\Phi_{\phi,r}\left(t\right)$ depends entirely on the orbital evolution parameters $\left\{p\left(t\right), e\left(t\right)\right\}$. Therefore, we still analyze the effects of the parameter $\zeta$ by numerically computing the relative phase deviation between cases with and without quantum gravity effects. Fig. \ref{fig_dPhi} shows the evolution of the dephasing $\left | \Delta \Phi \right | \equiv \left |{\Phi_{\phi}}_\mathrm{I, II}- {\Phi_{\phi}}_{\mathrm{Sch}}\right |$ over time for different $\zeta$ values in these two BH models, with both initial phase set to 0. 
The figure clearly shows that the phase difference grows with time for all chosen values of $\zeta$, and the departure from the Schwarzschild case becomes increasingly evident as $\zeta$ rises. This behavior indicates that the accumulated orbital phases may carry discernible imprints of quantum gravity effects in the GW signal. At fixed $\zeta$, the discrepancy is considerably stronger in the LQG-BH I spacetime, whereas the variation in the LQG-BH II background remains comparatively weak. These results suggest that LQG-BH I offers more favorable conditions for revealing quantum gravity signatures in EMRI waveforms, while the LQG-BH II scenario would require larger $\zeta$ to yield similarly measurable deviations.

\begin {figure}[h]
    \centering
     \vspace{0.35cm}
      \includegraphics[width = 1.03\linewidth] {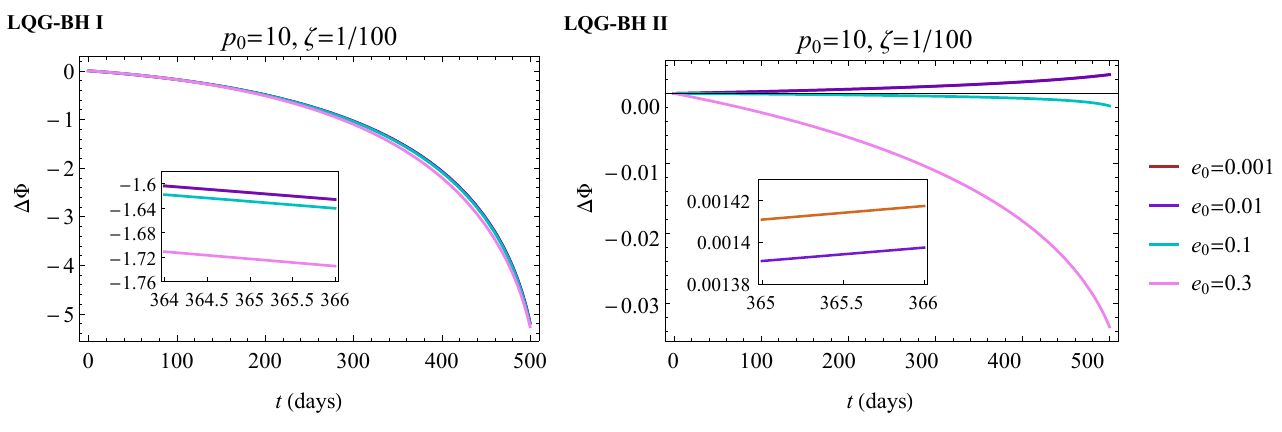}
    \caption {Time evolution of the dephasing $\Delta \Phi={\Phi_{\phi}}_\mathrm{I, II}- {\Phi_{\phi}}_{\mathrm{Sch}}$ for various initial eccentricities at a fixed quantum gravity value of $\zeta=1/100$, where the inset shows an enlarged view around one year of orbital evolution.
}
    \label{fig_dPhi_e0}
\end {figure}

It should be noted that Fig. \ref{fig_dPhi} displays the absolute phase deviation. The corresponding numerical values after one year of EMRI evolution are quantified in Table \ref{tab:dephasing}. In addition, Fig. \ref{fig_dPhi_e0} shows the signed dephasing (without taking the absolute value) as a function of time for fixed $\zeta=1/100$ and various initial eccentricities. Our analysis reveals that in the BH-I background, larger $\zeta$ generally induces a substantial phase delay relative to the Schwarzschild case, with little sensitivity to the initial eccentricity. For BH-II, however, the dephasing exhibits an intriguing eccentricity-dependent reversal: waveforms are advanced at low eccentricities but delayed at high eccentricities. These trends are qualitatively consistent with the findings of Ref. \cite{Chen:2025aqh} for quadrupolar waveforms, although the phase reversal in BH-II was not reported there. 
A more extended discussion is provided in Appendix \ref{Appendix}.

\begin{table}[h]
\centering
\caption{\centering Accumulated EMRI dephasing $\Delta\Phi$ over one year evolution for different initial eccentricities $e_0$ and quantum gravity parameters $\zeta$.}
\begin{tabular}{c|c|c|c}
\toprule[1pt]
\toprule[0.5pt]
\midrule
\ \quad $\quad e_{0} \quad$ \quad \  & \ \quad  $\quad \zeta \quad$ \quad \ &  \ \quad {\quad $\Delta\Phi_{\mathrm{I}}$ \quad} \quad \ &  \ \quad \quad {\quad $\Delta\Phi_{\mathrm{II}}$ \quad} \quad  \ \  \\
\midrule
\toprule[1pt]
\midrule
\multirow{4}{*}{0.01} & 1/1000 &  $-0.0161498$  &  $1.39096\times 10^{-5}$  \\
 & 1/100 &  $-1.61495$  &  0.00139095  \\ 
 & 1/10 &  $-161.263$  &  0.139062  \\
 & 4/10 &  $-2526.65$  &  2.21723  \\
\midrule
\toprule[0.5pt]
\midrule
\multirow{4}{*}{0.1} & 1/1000 &  $-0.0162933$  &  $-5.75953\times 10^{-6}$  \\
 & 1/100 &  $-1.6293$  &  $-0.000575956$  \\
 & 1/10 &  $-162.695$  & $-0.0576207$  \\
 & 4/10 &  $-2548.99$  &  $-0.927739$  \\
\bottomrule[1pt]
\end{tabular}
\label{tab:dephasing}
\end{table}


Thus far, the influence of $\zeta$ on the waveform phases has been assessed through relative dephasing. In the following section, we synthesize and study the full time-domain EMRI signals applicable to detection scenarios.

\section{Waveform generation and comparison}\label{Wave}

Several kludge waveform models have been proposed to efficiently generate EMRI waveforms. Recently, an improved AAK method, implemented within the FEW package, has been developed based on the original AAK model. The advantage of this method lies in its abandonment of frequency mapping, instead directly utilizing the inspiral trajectory as the foundation for waveform generation (for detailed technical information, please consult Refs. \cite{Katz:2021yft, Chua:2020stf}). The advanced AAK template greatly improves computational efficiency and is sufficient for qualitative studies of EMRI waveform characteristics.
In this section, we employ the FEW-based AAK method to construct EMRI waveforms with or without corrections involving the parameter $\zeta$. Specifically, the detectability of quantum gravity effects in GW signals is then systematically analyzed via the faithfulness approach under the LISA observational mission.

\subsection{Set up}

This subsection provides the requisite conditions and functional formulations for constructing EMRI gravitational waveforms, as a preparation for the upcoming LISA mission analysis.

To generate the FEW-based AAK waveforms, we employ the GW formalism developed in Ref. \cite{Barack:2003fp} within a time-dependent waveform frame. We define the unit vector $\hat{\mathbf{r}}$ that points from the detector to the source, while $\hat{\mathbf{L}}\left(t\right)$ specifies the direction of the secondary body's orbital angular momentum. 
Then, under the quadrupole approximation, the two polarization components of the GW strain field at the detector, i.e., the plus mode $h^{+}$ and the cross mode $h^{\times}$, can be written as the sum of a harmonic series over the orbital harmonics $n$:
\begin{eqnarray}
&&h^{+}=\sum_{n}{h^{+}_{n}} =\sum_{n} \left\{\left[1+ \left(\hat{\mathbf{r}} \cdot \hat{\mathbf{L}}\right)^{2}\right]\left( b_{n} \sin \left(2 \gamma\right)-a_{n} \cos \left(2 \gamma\right)\right) +\left[1-\left(\hat{\mathbf{r}} \cdot \hat{\mathbf{L}}\right)^{2}\right] c_{n}\right\}, \label{AAK_hp} \\[3mm]
&&h^{\times}=\sum_{n}{h^{\times}_{n}} =\sum_{n} 2\left(\hat{\mathbf{r}} \cdot \hat{\mathbf{L}}\right)\left( b_{n} \cos \left(2 \gamma\right) +a_{n} \sin \left(2 \gamma\right)\right) .
\label{AAK_hc}
\end{eqnarray}
The quantity $\gamma$ denotes the azimuthal angle of pericenter for an eccentric orbit. Within the equatorial-plane EMRI evolution considered in this work, it can be defined as $\gamma \equiv \Phi_{\phi}\left(t\right)-\Phi_{r}\left(t\right)$. 
The scalar product $\hat{\mathbf{r}}\cdot\hat{\mathbf{L}}$ is expressed as follows:
\begin{eqnarray}
\hat{\mathbf{r}} \cdot \hat{\mathbf{L}} =\cos \theta_{\mathrm{S}} \cos \theta_{\mathrm{L}}+\sin \theta_{\mathrm{S}} \sin \theta_{\mathrm{L}} \cos \left(\phi_{\mathrm{S}}-\phi_{\mathrm{L}}\right) ,
\end{eqnarray}
with $\left\{\theta_{\rm{S}},\phi_{\rm{S}}\right\}$ and $\left\{\theta_{\rm{L}}, \phi_{\rm{L}}\right\}$ denoting the ecliptic latitude and longitude of the source position and the orbital angular momentum direction, respectively \cite{Apostolatos:1994mx}. 
In addition, the constituents $\{a_n, b_n, c_n\}$ are formulated by Peters-Mathews type \cite{Peters:1963ux, Peters:1964zz}: 
\begin{eqnarray}
a_{n} &=& -n {A}\left(J_{n-2}\left(n e\right) - 2 e J_{n-1}\left(n e\right) + 2 J_{n}\left(n e\right)/ n + 2 e J_{n+1}\left(n e\right)  \right. \nonumber\\[3mm]
&& \left. -J_{n+2}\left(n e\right)\right) \cos \left(n \Phi_{r}(t)\right), \nonumber\\[3mm]
b_{n} &=& -n {A}\left(1-e^{2}\right)^{1 / 2}\left(J_{n-2}\left(n e\right) - 2 J_{n}\left(n e\right)+J_{n+2}\left(n e\right)\right) \sin \left(n \Phi_{r}(t)\right), \label{eq:abc} \\[3mm]
c_{n} &=& 2 {A} J_{n}\left(n e\right) \cos \left(n \Phi_{r}(t)\right) . \nonumber
\end{eqnarray}
In this expression, the sympol $J$ represents the Bessel function of the first kind. The amplitude coefficient is given by ${A}=(M\Omega_\phi)^{{2}/{3}}m/D_{\rm{L}}$, where $D_{\rm{L}}$ is the luminosity distance to the source \cite{Markovic:1993cr}.

\begin {figure}[h]
    \centering
     \vspace{0.35cm}
      \includegraphics[width = 1.03\linewidth] {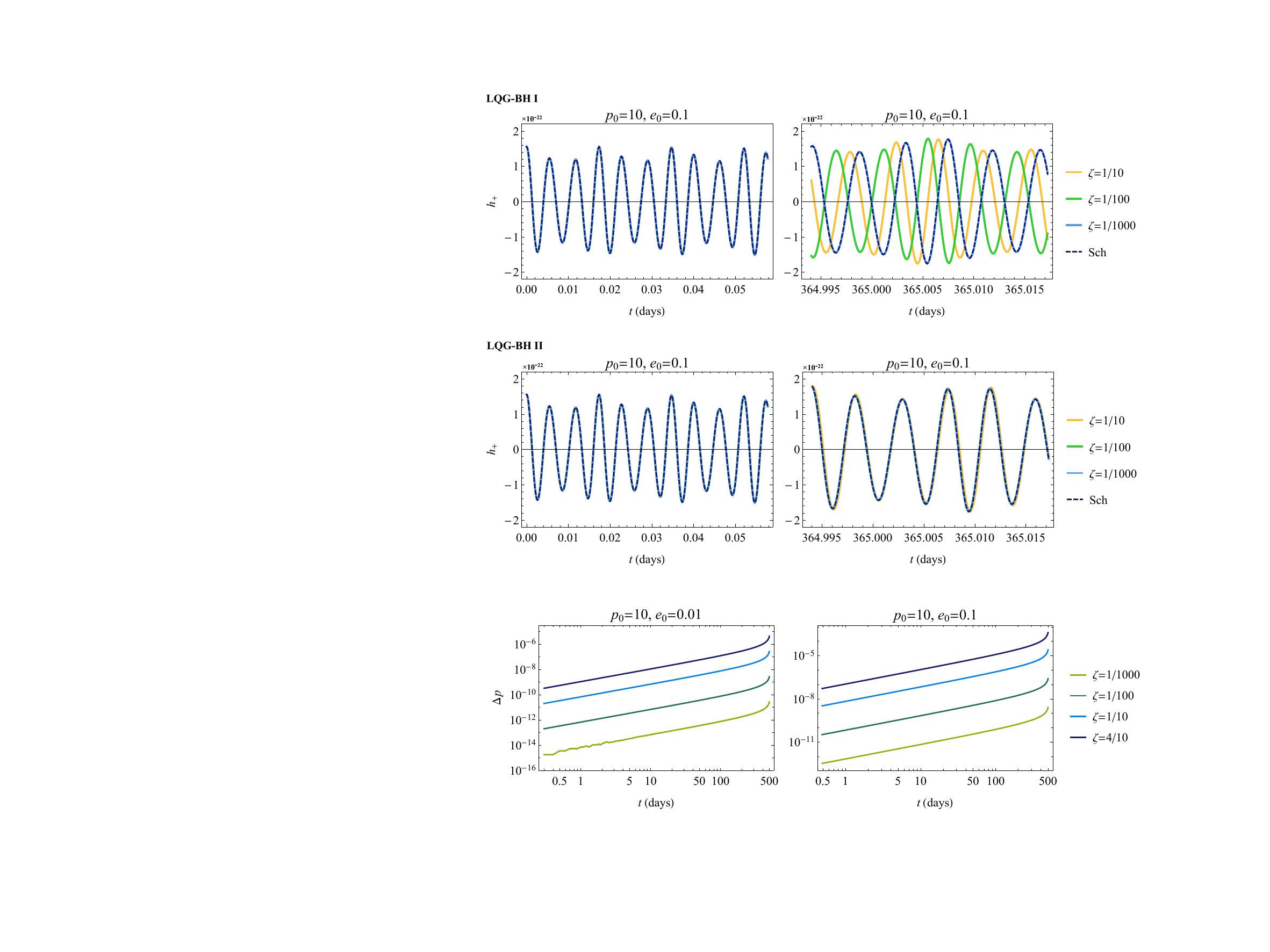}
    \caption {The $h^{+}$ polarization of EMRI waveforms for different values of the quantum gravity parameter $\zeta$
.}
    \label{fig_WF_e01f10}
\end {figure}

Using the numerical framework established above, we generate the $h^{+}$ polarization mode of EMRI waveforms for the Schwarzschild background and for LQG-BH backgrounds with various values of $\zeta$, assuming an initial eccentricity of 0.1 and a mass ratio of $10^{-5}$. Apart from these parameters, the remaining initial parameters required for waveform generation are summarized in Table \ref{tab:value}. 
The resulting signals at the initial moment and after one year of evolution are presented in Fig. \ref{fig_WF_e01f10}, where solid colored curves represent the quantum-corrected cases and dashed curves indicate the Schwarzschild reference.
The left panel reveals that at the initial time, the waveforms in both BH-I and BH-II geometries are nearly indistinguishable from the Schwarzschild result, regardless of $\zeta$ values. After one year of evolution, the BH-I background with larger $\zeta$ exhibits clear phase separation from the Schwarzschild case, whereas the deviation remains negligible for $\zeta = 1/1000$ (upper-right panel). By contrast, in the BH-II scenario, even comparatively large $\zeta$ values produce only minimal departures (bottom-right panel). These observational trends are consistent with the phase-difference analysis detailed in subsection \ref{evolution}.

These findings further suggest that cumulative phase evolution in the BH-I background enhances sensitivity to possible quantum gravity effects. In other words, GW observations of EMRIs in such spacetimes could produce signals distinct from those in the Schwarzschild case, while in the BH-II background, although the parameter $\zeta$ influences the waveforms, its impact is weak and likely beyond observational detectability.

\begin{table}[th]
\centering
\caption{\centering Numerical values of corresponding parameters for EMRI waveforms generation.}
\vspace{0.5em}
\begin{tabular}{c|c|c|c}
\toprule[1pt]
\toprule[0.5pt]
\midrule
\quad {Intrinsic Parameters} \ \ &  \quad {Values } \quad & \quad {Extrinsic Parameters } \quad  & \quad \quad {Values} \quad \quad  \\
\midrule
\toprule[1pt]
\midrule
\quad $M  \quad$ & \quad  $10^6\,M_{\odot}  \ $  & \quad $D_{\rm{L}}\quad$ & \quad 1 Gpc \quad  \\
\quad $m  \quad$  & \quad  $10\,M_{\odot}$  \ &  \quad $\theta_{\rm{S}}\quad$  & \quad $\pi/3$ \quad  \\
\quad $p_{0} \quad$  &  \ \ 10   \ \ & \quad  $\phi_{\rm{S}}\quad$   & \quad  $\pi/2$ \quad    \\
\quad $e_{0}  \quad$   &  \ \ 0.01, 0.1 \ \ &  \quad $\theta_{\rm{L}}\quad$  & \quad $\pi/4$ \quad  \\
\quad ${\Phi_{\phi}}_{0}  \quad$ & \ \  0 \ \  &  \quad $\phi_{\rm{L}}\quad$  & \quad $\pi/4$ \quad \\
\quad ${\Phi_{r}}_{0}  \quad$ & \ \  0 \ \ & \ \ & \ \ \\
\quad $\zeta \quad$ \quad & \ 1/1000, 1/100, 1/10 \quad & \ \ & \ \ \\
\midrule
\bottomrule[1pt]
\end{tabular}
\label{tab:value}
\end{table}

\subsection{Data analysis with faithfulness}

In this subsection, we carry out a faithfulness analysis within the LISA framework to quantify how quantum gravity effects imprint on detectable EMRI waveforms. The analysis directly compares waveforms from LQG-corrected and Schwarzschild backgrounds, thereby assessing LISA’s potential to identify such deviations.

LISA \cite{Cutler:1997ta, Barack:2003fp} is a space-based GW observatory designed to detect low-frequency gravitational radiation by monitoring the phase or frequency shifts of laser beams exchanged among three widely separated spacecraft. The constellation’s barycenter moves on a circular heliocentric orbit trailing the Earth by roughly $20^{\circ}$, and the plane defined by the spacecraft is tilted by $60^{\circ}$ relative to the ecliptic, a geometry that preserves an essentially equilateral configuration throughout the mission. This triangular array forms three long baselines, which can be combined into an effective pair of two-arm interferometers, allowing simultaneous measurement of the two GW polarizations. The corresponding strain amplitudes for these two detectors are denoted as $h_{\rm{d1}}\left(t\right)$ and $h_{\rm{d2}}\left(t\right)$. In the regime where the gravitational wavelength is much larger than the interferometer arm length, $h_{\rm{d1}, \rm{d2}}\left(t\right)$ can be expressed in terms of a sum over the $n$-th harmonic contributions of the two polarization components: $h_{\rm{d1}, \rm{d2}}\left(t\right)= \sum_{n} h_{\left\{\rm{d1}, \rm{d2}\right\},\,n}\left(t\right)$, where the response function
\begin{eqnarray}
h_{\left\{\rm{d1}, \rm{d2}\right\},\,n}\left(t\right)= {\sqrt{3}}/{2}\left( h_{n}^{+}\left(t\right){F_{\rm{d1}, \rm{d2}}^{+}}\left(t\right)+ h_{n}^{\times}\left(t\right){F_{\rm{d1}, \rm{d2}}^\times} \left(t\right)\right) .
\label{LISA_hp}
\end{eqnarray}
On the right-hand side of the above expression, the coefficient ${\sqrt{3}}/{2}$ is a scale factor arising from the $60^{\circ}$ opening angle between LISA's arms. The waveform components $h_{n}^{+}\left(t\right)$ and $h_{n}^{\times}\left(t\right)$ are already defined in Eqs. \eqref{AAK_hp} and \eqref{AAK_hc}, ${F_{\rm{d1}, \rm{d2}}^{+}}\left(t\right)$ and ${F_{\rm{d1}, \rm{d2}}^\times}\left(t\right)$ are the interferometer beam-pattern functions, which depend on the trigonometric combinations of the source location $\left\{\theta_{\rm{S}},\phi_{\rm{S}}\right\}$ and the orbital angular momentum direction $\left\{\theta_{\rm{L}}, \phi_{\rm{L}}\right\}$ in the ecliptic coordinate system \cite{Barack:2003fp, Cutler:1997ta, Apostolatos:1994mx}. It is important to note that, since this work is restricted to non-rotating BH spacetimes, the precession of $\hat{\mathbf{L}}\left(t\right)$ induced by the central BH's spin is neglected, making it appropriate to treat $\left\{\theta_{\rm{L}}, \phi_{\rm{L}}\right\}$ as fixed parameters.

Furthermore, in modeling the GW response of LISA, we incorporate the Doppler phase modulation generated by the detectors' annual motion around the Sun, which becomes increasingly relevant for long observation times. This modulation is introduced in Eq. \eqref{eq:abc} by the replacement: $n\Phi_{r}\left(t\right) \to n\left(\Phi_{r}\left(t\right) + \Phi_{\rm{D}}\left(t\right) \right)$, the Doppler phase $\Phi_{\rm{D}}\left(t\right)$ is defined as \cite{Cutler:1997ta, Barack:2003fp}
\begin{eqnarray}
\Phi_{\rm{D}}\left(t\right) \equiv 2\pi f_{n}\left(t\right) R\sin\theta_{\rm{S}}\cos\left(2\pi t/T-\phi_{\rm{S}} \right)  
\end{eqnarray}
with $R=1\,\mathrm{AU}$, $T=1\,\mathrm{year}$, and $f_{n}\left(t\right)$ denoting the GW frequency associated with $n$-harmonic of the orbital frequencies.

To quantitatively assess the detection threshold of LISA for EMRI signals containing quantum gravity corrections, we carry out a systematic comparison of waveform overlaps between Schwarzschild and the two quantum-corrected models (BH-I and BH-II). A central measure of signal detectability is the SNR. In our simulations, the waveform is injected into noise and rescaled to attain a prescribed optimal SNR, defined by the standard relation \cite{Cutler:1994ys}: $\mathrm{SNR}\equiv\rho=\sqrt{\left \langle h|h \right \rangle}$, where $h$ denotes the EMRI strain template and $\left \langle \,|\, \right \rangle $ represents the noise-weighted inner product in the frequency domain. The statistical properties of the detector noise define a natural inner product on the signal space. When evaluating the similarity between two signals, such as LQG-corrected waveform $h_{a}\left(t\right)$ and its Schwarzschild counterpart $h_{b}\left(t\right)$, this inner product is given by
\begin{eqnarray}
\left \langle h_{a}|h_{b}\right \rangle = 2 \int_{f_{\min}}^{f_{\max}} \frac{\tilde{h}_{a}^{*}\left(f\right)\,\tilde{h}_{b}\left(f\right) + \tilde{h}_{a}\left(f\right)\,\tilde{h}_{b}^{*}\left(f\right)} {S_n\left(f\right)} \, \mathrm{d}f . 
\label{innerp}
\end{eqnarray}
From the above expression, the tildes indicate Fourier transforms of the strain signals, followed by the agreement $\tilde{h}\left(f\right)= \int_{-\infty}^{+\infty} h\left(t\right) e^{2\pi i f t}\mathrm{d}t$ \cite{Finn:1992wt}. The asterisk denotes complex conjugation, such that $h\left(f\right)=h^{*}\left(-f\right)$, and the frequency limits $\left\{f_{\mathrm{max}}, f_{\mathrm{min}}\right\}$ correspond to LISA's low-frequency cutoff ($10^{-4}$ Hz) and the orbital frequency after one year of inspiral. 
The function $S_{n}\left(f\right)$ is the one-sided noise power spectral density of LISA. In this work, we adopt the Robson-Cornish model \cite{Robson:2018ifk}, which includes both instrumental noise and the confusion noise generated by unresolved Galactic binaries. This inner product naturally weights each frequency component by the detector sensitivity and forms the basis of matched filtering, SNR estimation, and likelihood-based inference in GW data analysis.

Next, using identical physical parameters, we define the faithfulness $\mathcal{F}$ between two waveform templates as:
\begin{eqnarray}
\mathcal{F}\left(h_{a},\,h_{b}\right) \equiv \max_{t_{c},\phi_{c}} \frac{\left \langle h_{a}|h_{b} \right \rangle }{\sqrt{\left \langle h_{a}|h_{a} \right \rangle \left \langle h_{b}|h_{b} \right \rangle } } ,
\end{eqnarray}
where $\max_{t_{c},\phi_{c}}$ denotes maximization over the time and phase offsets \cite{Lindblom:2008cm}. The faithfulness serves as an effective statistical measure for assessing the suitability of waveform models in parameter estimation. In particular, this metric enables us to quantify both the impact of the correction parameters on the waveform and the degree of consistency between a template and the true signal. By construction, the value of $\mathcal{F}$ lies in range $\left[0,\,1\right]$, with $\mathcal{F}=1$ corresponds to identical waveforms, $\mathcal{F}=0$ to zero correlation. 
For a signal with SNR $\rho$, statistical fluctuations in the Fisher matrix induce a characteristic waveform mismatch of order $D/(2\rho^2)$ in a $D$-dimensional intrinsic parameter space \cite{Lindblom:2008cm}. Quantum gravity effects can be distinguished by LISA only if the faithfulness between the two waveforms satisfies $\mathcal{F} \lesssim 1 - D/(2\rho^2)$, i.e., when the mismatch exceeds the typical statistical fluctuations.
Based on the number of intrinsic parameters and initial numerical settings in this work (see Table \ref{tab:value}), we assume that for an EMRI signal with $\rho=30$, LISA can distinguish Schwarzschild from LQG-modified waveforms when the faithfulness $\mathcal{F}$ drops below the threshold 0.996 \cite{Chatziioannou:2017tdw, Mangiagli:2018kpu, Maselli:2021men, Barsanti:2022vvl}.

\begin {figure}[h]
    \centering
     \vspace{0.35cm}
      \includegraphics[width = 1.03\linewidth] {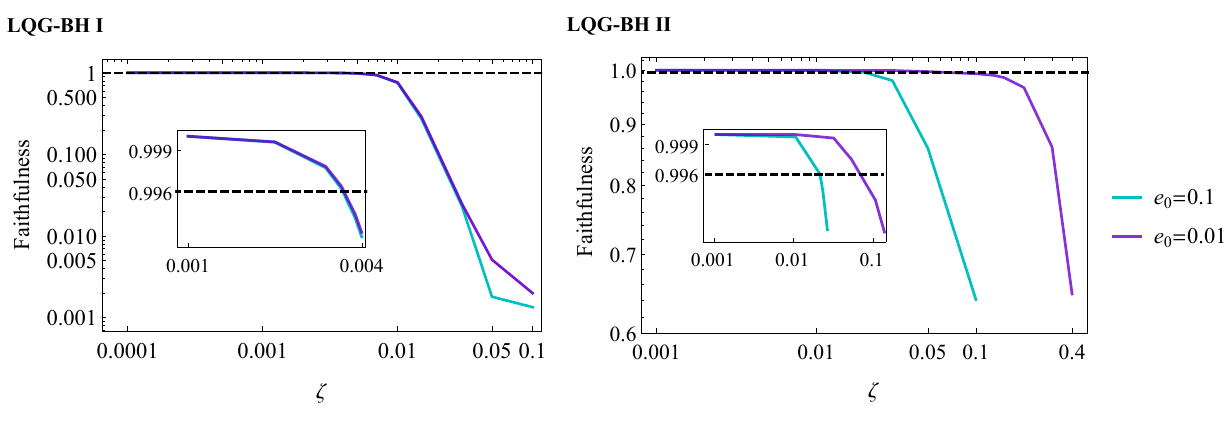}
    \caption {Faithfulness $\mathcal{F}$ between EMRI waveforms in the LQG and Schwarzschild BH backgrounds
. The inset shows an enlarged view near the threshold 0.996.}
    \label{fig_FFS_0.996}
\end {figure}

The numerical results for the faithfulness analysis of the $h^{+}$ polarization waveforms are presented in Fig. \ref{fig_FFS_0.996}, as a function of the parameter $\zeta$ for a one-year observation period. We examine cases with initial eccentricities $e_0 = 0.01$ and $0.1$. Table \ref{tab:constrain} summarizes the values of $\zeta$ at which the faithfulness equals the threshold $\mathcal{F}_{\mathrm{thr}}=0.996$, indicated by the black dashed line in Fig. \ref{fig_FFS_0.996}.

\begin{table}[th]
\centering
\caption{\centering Lower limits on the LQG parameter $\zeta$ detectable by LISA for different initial eccentricities $e_0$.}
\vspace{0.5em}
\begin{tabular}{c|c|c}
\toprule[1pt]
\toprule[0.5pt]
\midrule
\multirow{3}{*}{\quad Type-I BH \quad} \quad & \quad $ e_{0} \quad$ & \quad $ \zeta \quad$   \\
\cline{2-3}
& \quad $0.1 \quad$ & \quad $ 0.003452 \quad$   \\
& \quad $0.01 \quad$ & \quad $ 0.003492 \quad$   \\
\midrule
\toprule[1pt]
\midrule
\multirow{3}{*}{\quad Type-II BH \quad} \quad & \quad $ e_{0} \quad$ & \quad $ \zeta \quad$   \\
\cline{2-3}
& \quad $0.1 \quad$ & \quad $ 0.02008 \quad$   \\
& \quad $0.01 \quad$ & \quad $ 0.05768 \quad$   \\
\midrule
\bottomrule[0.5pt]
\bottomrule[1pt]
\end{tabular}
\label{tab:constrain}
\end{table}

In the left panel of Fig. \ref{fig_FFS_0.996}, comparing the faithfulness of LQG-BH I and Schwarzschild waveforms, the two curves maintain close alignment even after one year, deviating appreciably only for larger values of $\zeta$. As discussed previously, this behavior arises from the weak sensitivity of the LQG-BH I model to the initial eccentricity, implying that for small $\zeta$ the LQG-I-corrected EMRI signals at different evolutionary stages may difficult to tell apart. Simultaneously, the signal overlap declines precipitously with increasing $\zeta$, falling below the detection threshold already at $\zeta\approx 0.0035$, and by $\zeta = 1/10$, the faithfulness $\mathcal{F}$ drops to approximately $0.001$. 
Under the same configuration, the comparison between LQG-BH II and Schwarzschild (right panel of Fig. \ref{fig_FFS_0.996}) shows a similar trend, but with systematically higher overlap. Notably, the curves exhibit distinct separation as $\zeta$ increases, rendering the two waveforms distinguishable for $\zeta \gtrsim 0.02$ in the mass ratio $10^{-5}$ system. Furthermore, at a fixed $\zeta$, higher eccentricity drives the faithfulness below the detection threshold earlier, and vice versa. Consequently, higher initial eccentricity enhances deviations in the EMRI waveform, enabling clearer identification of quantum gravity features and improving LISA's ability to discriminate such effects.

The present analysis highlights three key points:
(i) quantum gravity corrections induce dephasing in the EMRI signal;
(ii) this phase shift leads to a mismatch between classical templates and waveforms with nonzero $\zeta$, potentially biasing parameter estimation; and
(iii) for identical setups, the parameter $\zeta$ exerts a more pronounced influence in the BH-I background compared to BH-II.
Taken together, these results indicate that LISA's one-year observation of EMRI events may be sensitive to quantum gravity signatures as small as $\zeta\gtrsim 10^{-3}$ for BH-I or $\zeta\gtrsim 10^{-2}$ for BH-II model.

\section{Conclusions and discussions}\label{conclusion}

Within the framework of the Hamiltonian constraint, C. Zhang et al. established effective physical metrics by imposing general covariance conditions, yielding two static, spherically symmetric LQG‑BH solutions under distinct quantization schemes. Both models introduce quantum gravity modifications parameterized by $\zeta$.
In this work, we have investigated the imprint of these subtle quantum corrections on the exterior spacetime by employing EMRIs as high-precision probes.
Using the efficient semi-classical kludge approach, we constructed waveform templates based on an improved AAK model within the FEW package, comparing them with the corresponding signals in a classical Schwarzschild spacetime. We assessed the ability of future space-borne detectors such as LISA to discriminate the deviations and derived observational constraints on the parameter $\zeta$. Throughout the analysis, similarities and differences between the two LQG-BH backgrounds have been systematically examined.

Our numerical results show that although $\zeta$ enters only at subleading or higher PN orders, its effect accumulates over time through cumulative dephasing, leading to a non‑negligible deviation that grows with $\zeta$. A clear contrast emerges between the two models: in LQG-BH I, $\zeta$ induces a significant phase lag largely insensitive to initial eccentricity, whereas in LQG-BH II, the deviation is subtler, manifesting as a small phase lead at low eccentricities and transitioning to a lag for higher eccentricities. These distinctive features are consistently reflected in the $h^{+}$ polarization waveforms of the corresponding AAK EMRI signals, implying that such signatures can be imprinted within the gravitational signal and may serve as viable templates for observational tests.

Based on the faithfulness analysis with SNR $\rho=30$ and the detection threshold $\mathcal{F}_{\mathrm{thr}}=0.996$, we find that for an EMRI with mass ratio $10^{-5}$, LISA can detect deviations at the level of $\zeta\gtrsim10^{-3}$ in the BH-I spacetime for initial eccentricities of 0.01 and 0.1, while the corresponding sensitivity in the BH-II model is $\zeta\gtrsim10^{-2}$. Thus, the BH-I background offers a higher probability of detecting quantum gravity effects. Moreover, for a fixed value of $\zeta$, larger orbital eccentricities enhance the prospects of detecting such signatures in space-based EMRI detections.

These results indicate that EMRIs offer a new and effective avenue for constraining the parameter $\zeta$, yielding constraints significantly tighter than those from other strong-field probes such as BH shadows (M 87: $\zeta \gtrsim2.30$; Sgr A$^{*}$: $\zeta \gtrsim2.87$) \cite{Shu:2024tut, Konoplya:2024lch} and the periapsis advance of the S2 star ($\zeta \gtrsim0.74$) \cite{Xamidov:2025oqx}. Our bounds are complementary to existing limits and suggest that future astrophysical observations may provide further evidence for, or constraints on, potential quantum gravity effects.

We emphasize that the present analysis relies on the Peters-Mathews quadrupole approximation, and therefore the resulting constraints on $\zeta$ should be regarded as order of magnitude estimates. More reliable bounds or theoretical predictions, such as those incorporating complete higher PN theory or high precision numerical waveform methods, are left for future work.

\acknowledgments

R.T. Chen is grateful to Yunlong Liu and Prof. Andrea Maselli for their valuable discussions and clarification of related questions. This work is supported by National Key R$\&$D Program of China (No. 2023YFC2206703), the Natural Science Foundation of China under Grant Nos. 12375055, 12505078 and 12505085, the Jiangsu Postgraduate Research and Practice Innovation Program under Grant No. KYCX25$_{-}$3925, the China Postdoctoral Science Foundation (No. 2025T180931), and the Jiangsu Funding Program for Excellent Postdoctoral Talent (No. 2025ZB705).

\appendix

\section{Eccentricity-dependent dephasing in LQG-BH II}\label{Appendix}

In subsection \ref{evolution}, we have investigated the relative dephasing in LQG-BH spacetimes during the inspiral evolution within EMRI. From the results of LQG-BH II, we find an interesting behavior of the waveform phase when varying the initial eccentricity while keeping the parameter $\zeta=1/100$ fixed. 

\begin {figure}[th]
    \centering
     \vspace{0.35cm}
      \includegraphics[width = 0.8\linewidth] {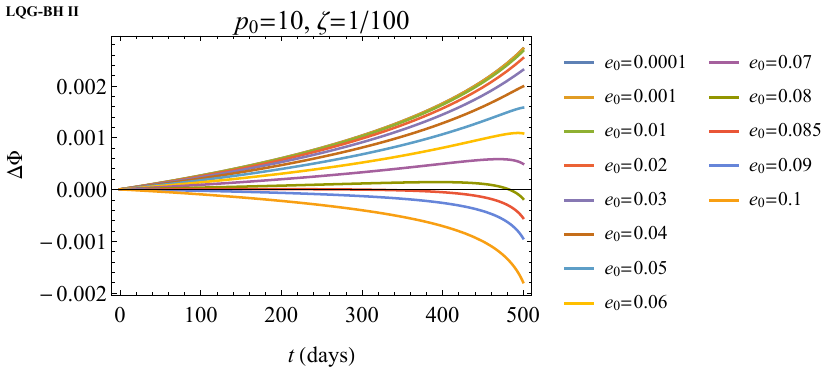}
    \caption {The orbital dephasing $\Delta \Phi={\Phi_{\phi}}_\mathrm{II}-{\Phi_{\phi}}_{\mathrm{Sch}}$ as function of the evolution time for various initial eccentricities. The fixed initial parameters $\left\{M, m, p_0, \zeta, {\Phi_{\phi}}_0, {\Phi_{r}}_0\right\}=\left\{10^{6} M_{\odot}, 10 M_{\odot}, 10, 1/100, 0, 0\right\}$ correspondingly.}
    \label{fig_dPhi2_multie0}
\end {figure}

To illustrate this, we show in Fig. \ref{fig_dPhi2_multie0} the dephasing $\Delta\Phi$ for a range of initial eccentricities. It is evident that for small eccentricities, quantum gravity corrections lead to a phase advance relative to the Schwarzschild waveform during the orbital evolution. As the eccentricity increases, this trend reverses, and the phase shift becomes a delay that grows with larger $e_{0}$. The transition is particularly evident for approximately $e_{0}=0.08$ and above. As discussed in Refs. \cite{Bonga:2019ycj, Barsanti:2022vvl}, the phase difference $\Delta \Phi \gtrsim 0.1$ radian is considered resolvable for a GW signal with $\mathrm{SNR}\approx 30$, which implies that such characteristics may serve as distinctive signatures in waveform templates for GW observations sensitive to quantum gravity effects for larger $\zeta$ in BH-II.

\bibliographystyle{utphys}
\bibliography{Ref}
\end{document}